# FIPA agent based network distributed control system

V. Gyurjyan, D. Abbott, G. Heyes, E. Jastrzembski, C. Timmer, E. Wolin
*TJNAF, Newport News, VA 23606, USA*

A control system with the capabilities to combine heterogeneous control systems or processes into a uniform homogeneous environment is discussed. This dynamically extensible system is an example of the software system at the agent level of abstraction. This level of abstraction considers agents as atomic entities that communicate to implement the functionality of the control system. Agents' engineering aspects are addressed by adopting the domain independent software standard, formulated by FIPA. Jade core Java classes are used as a FIPA specification implementation. A special, lightweight, XML RDFS based, control oriented, ontology markup language is developed to standardize the description of the arbitrary control system data processor. Control processes, described in this language, are integrated into the global system at runtime, without actual programming. Fault tolerance and recovery issues are also addressed.

## 1. INTRODUCTION

Over time controls in high energy and nuclear physics experiments are inevitably becoming more heterogeneous and diverse. Industrial process control software and hardware growth should not be underestimated. More and more we see that industrial proprietary control systems with their specific hardware are being used in physics experiments, and that systems are robust and reliable. The use of PC's in control systems is widely supported since they are inexpensive and extensible. Even PC based control systems developed in the Microsoft windows environment are used. We think that the control and data acquisition systems for future experiments will face problems in areas such as interoperability, scalability, and standard user interface. A central concern of this research work is to address those issues by developing an open architecture control system. An open architecture design will add flexibility by shifting the focus of the control systems from being hardware-centric to more software-centric.

The goal of this project is to develop a multi-agent control system containing distributed control and application entities that collaborate dynamically to satisfy control objectives of physics experiment. Agents are capable of addressing both knowledge processing and control specific actions simultaneously in a real-time distributed environment. The proposed system will integrate data acquisition and the slow control components of the experiment in the homogeneous multi-agent environment. The system is required to cope also with the dynamic reconfiguration of the control environment in real time by auto-generating or specializing specific control agents whenever new controls are added, or control relationships are changed. In this paper we present the work in progress to design and develop an agent-based network-distributed control system.

The following section will introduce agent technologies and agent specification standards used in this project. Sections three and four will describe control agent platform architecture and specialized agents' descriptions and behaviors. Section five will discuss the developed metadata used to describe control systems and processes planned to be integrated into the global control environment. Finally, we conclude with a brief discussion of our future plans.

## 2. SOFTWARE AGENTS

An agent is a software entity capable of acting intelligently on behalf of a user, in order to accomplish a given task. Agents, like humans, co-operate so that a society of agents can combine their efforts to achieve a desired goal. The characteristic properties of the agents are:

- autonomy,
- proactive intelligence (agents do not simply to act in response to their environment, but are able to take initiative),
- temporal continuity (they are continuously running processes),
- mobility,
- rationality/benevolence (agents don't have conflicting goals), and
- adaptive intelligence (agents have the ability to learn).

Compared to OO objects, software agents have their own thread of control, localizing not only code and state but their invocations as well. In other words, agents themselves define when and how to act.

In an open and distributed, agent-based, integrated control environment, the need of standard mechanisms and specifications are vital for ensuring interoperability of the autonomous agents. At the very minimum we need:

- a common agreed means by which agents can communicate with each other so that they can exchange information, delegate control tasks, etc.
- facilities whereby agents can locate each other.
- a unique way of agent identification.
- a means of interacting with users.
- a means of migrating from one platform to another.

Foundation for Intelligent Physical Agents (FIPA) is the most promising standardization effort in the software agent world [1]. The FIPA agent reference model was chosen to provide the normative framework within which agents can be deployed and operate. FIPA specification establishes the logical reference model for the creation, registration, location, communication, migration and retirement of agents.

The FIPA standard do not attempt to prescribe the internal architecture of agents nor how they should be implemented, but it specifies the interface necessary to support interoperability between agent systems. THE FIPA agent platform (AP) suggests the following mandatory components or normative agents.

Directory facilitator (DF) provides "yellow pages" services to other agents. Agents may register their services with the DF or query the DF for information on other agents. An agent platform can have multiple DF's thus providing the possibility of creating communities or domains of agents. DF's can register with each other forming a federation of domains.

An agent management system (AMS), provides agent name services ("white pages"), and maintains an index of all agents which currently are registered with an AP. AMS exerts supervisory control over access to and the use of an AP. This normative agent is responsible for creation, deletion, and migration of the agents.

An agent communication channel (ACC) is the message transport system which controls all the exchanges of messages within the platform, as well as messages to/from remote platforms.

From the variety of the FIPA specification implementations (ZEUS, GRASSHOPPER, MOLE, RETSINA, FIPA-OS, etc.) we chose JADE (Java Agent Development Framework). JADE simplifies agent-based application development while ensuring standard compliance through a comprehensive set of FIPA services and FIPA agents [2].

## 3. DESIGN CRITERIA FOR THE SYSTEM

The overall goal is to create a real time, auto-extensible system, where completely heterogeneous processes can coexist and communicate with one another as peers, simplifying organization of the algorithmic controls and feedback mechanisms to achieve safe and efficient control of the experiment. To achieve this goal we should be able to describe individual control processes within the system, which will in turn encapsulate non-agent software components and allow already existing agents to communicate with those foreign processes as familiar agents. Figure 1 shows the main components of the design architecture.

Special control specific meta-data were developed using RDFS (Resource Definition Framework Schema). Meta-data then were imported into the Protégé-2000. Protégé-2000 is an integrated software tool able to develop and construct a knowledge base, and was developed at Stanford University [3]. The fact that Protégé-2000 is highly customizable makes it a perfect editing environment for new, imported semantic languages. So, created control semantic language meta-classes, available in Protégé-2000 are used to describe the semantics of the particular control process.

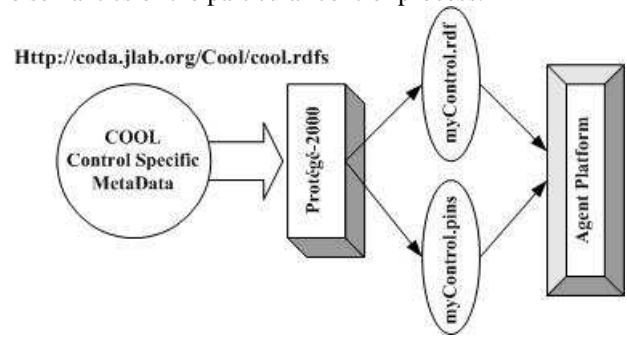

Figure 1: Design architecture of the system.

The Protégé-2000 user will construct a control specific knowledge base. The created control knowledge base, saved in the RDF format or Protégé-2000 text output (pins), will be picked up by the specialized high-level mediator agents to generate wrapper agents for the described processes.

## 4. AGENT PLATFORM ARCHITECTURE

The control agent platform includes FIPA specified mandatory agents (ACC, AMS and DF), provided by Jade. All agent communication is performed through message transfer. Message representation is based on the Agent Communication Language (ACL) formulated by FIPA [4]. ACL is a language with well-defined syntax, semantics and pragmatics, based on speech act theory, containing two distinct parts:

- the communicative act
and
- the content of the message.

Communicative acts have a precise, declarative meaning independent of the content of a massage, and will extend any intrinsic meaning that the message content itself may have. The software architecture of the system is based on the coexistence of several Java Virtual Machines (JVM), communicating with each other through Java RMI (Remote Method Invocation). Software agents are grouped into virtual clusters or domains with their specialized functionalities. Control agents in the same domain share the single JVM, which plays a role of a basic agent container. An agent container provides a complete run time environment for agent execution and allows agents to concurrently execute on the same host. Each agent container is a multithreaded execution environment composed of one thread for every agent. Control agents are active objects, having more then one behavior and can engage in multiple, simultaneous activities. In order to minimize the number of threads required to run the agent platform, each control agent implements a scheduler, which carries out a round-robin non-preemptive policy among all behaviors registered with the agent. Agent behaviors can be added and removed at run time. Figure 2 shows the architecture for the control system agent platform.

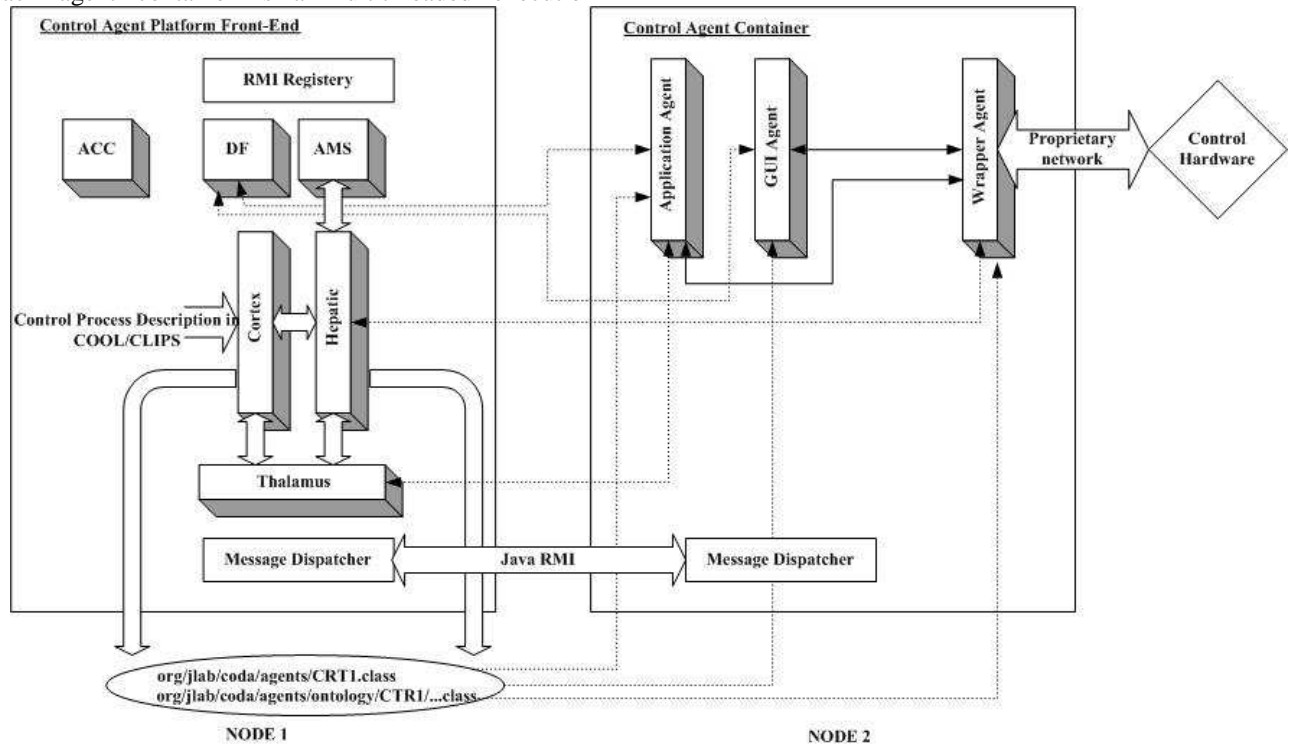

Figure 2: Control system agent platform.

The architectural structure of the system is based on hierarchies of the agent containers dispersed over the network. Control specific agents are dynamically created and grouped into virtual clusters. Control agents, virtual clusters as well as agent containers, can be created and destroyed as needed within the control platform.

The Front-End is a special container running the FIPA normative agents and developed high level mediator agents, taking care of control agents' management and overall coordination of the system. The Front-End container also maintains an RMI registry internally, used by other agent containers to register them with Front-End and to join the control agent platform.

The Hepatic high level mediator agent is responsible for the agent platform management and recovery processes. This agent is in charge of creation, recovery and removal of the agents, agent clusters or entire containers. The Hepatic agent will also resurrect any crashed agents or containers, thus achieving agent platform stability and fault tolerance.

Overall coordination of the data acquisition agent cluster is accomplished by the Thalamus agent, which ensures that partial, local solutions to control problems are integrated into the global framework. The Cortex agent is the interface between a specific control system designer and the global control system. It enables the integration of the controls provided by non-agent software into a multi-agent control community. The Cortex will auto-generate a wrapper agent, as well as all necessary ontology classes by parsing the specific control system description written in the developed Control Oriented Ontology Language (COOL). Agents on the control platform can relay messages to the wrapper agent and have them invoke action on the underlying proprietary hardware or software system.

## 5. CONTROL ORIENTED ONTOLOGY LANGUAGE AND CONTROL PROCESS DESCRIPTION

In order to embrace the heterogeneous nature of the control systems and information resources, we need to develop a language which will enable agents to understand domain specific services and devices. Using this language we can express information in the control system in a precise, machine-interpretable form, so the agents participating in the conversation will understand what the terms describing the specific control mean.

Light, extensible, control oriented ontology language (COOL) was developed to assist agents in sharing and annotating control specific information.

COOL uses RDFS (Resource Definition Framework Schema) developed by the WWW Consortium (W3C) [5] to define hierarchical descriptions of concepts (classes) in a control specific domain. COOL is extensible via XML-namespace and RDF based modularization. Properties of each class describe various features and attributes of the control concept. Logical statements describe relations among concepts. COOL also presents a set of predefined instances (for example instances for common used data types) to simplify knowledge base development. A node and arc diagram of the process description part of the COOL is illustrated in the Figure 3.

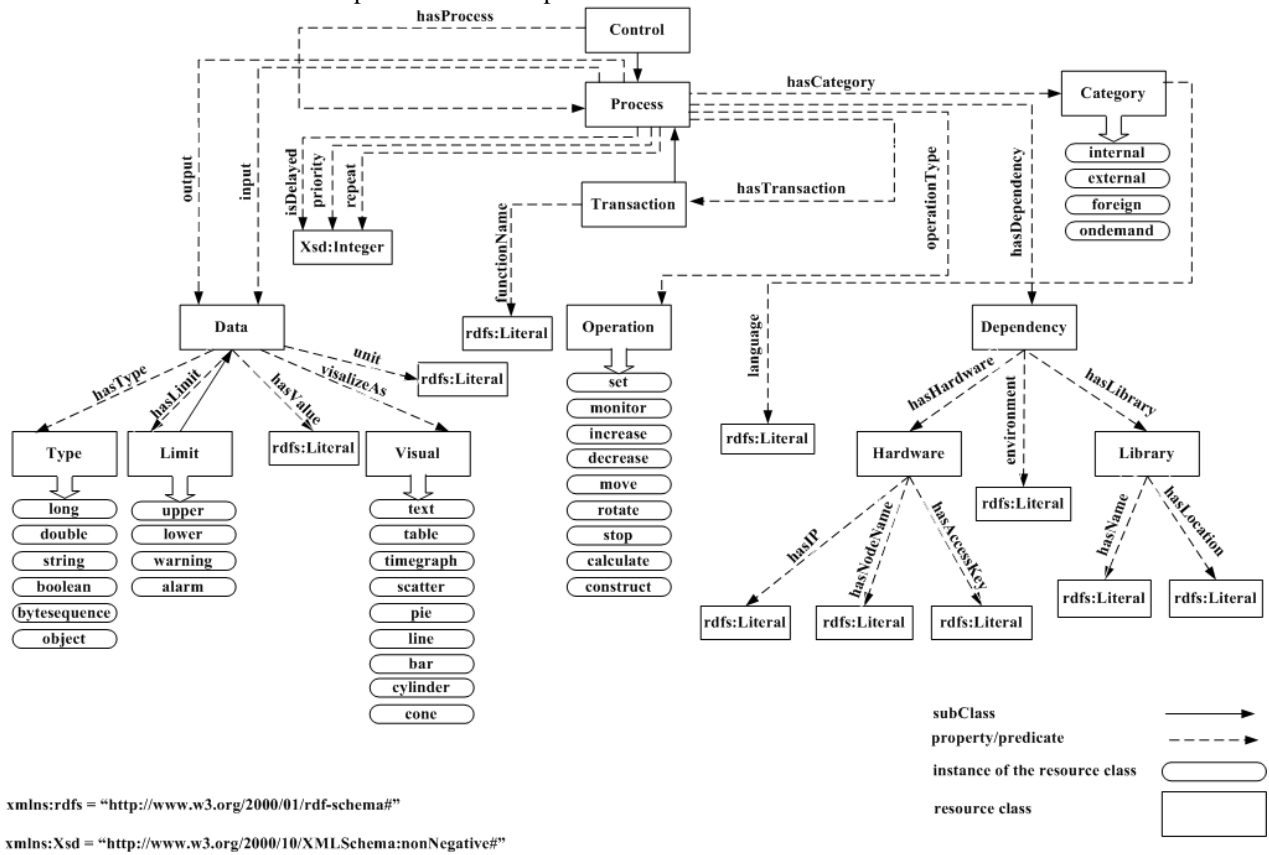

Figure 3: COOL schema and taxonomy.

A set of the classes to define graphical user interface components and their relations were also designed. Some of the Java swing classes are mirrored in COOL to simplify the design of the actual graphical interface for the specific control process.

Protégé-2000 was used to create a specific knowledge base (control system), by defining individual instances of the COOL classes, filling in specified slot value information and slot restrictions.

## 6. CONCLUSIONS

We proposed a new control system design based on software agent technology. It has been tested in a prototype based on FIPA compliant Jade agent platform.

Control process abstraction has been implemented through the newly developed control oriented ontology language, allowing description and integration into the general control environment not only of control processes but also of any processes in general.

High-level mediator agents specialized in agent platform management and system coordination were developed, increasing the entire control system reliability, robustness and fault tolerance.

External software or systems integration was accomplished by the special mediator agent, responsible for auto-generating wrapper agents on the platform, based on COOL or CLIPS (Protégé-2000 pins) description files.

We are still in the development stage of an integrated JLAB data acquisition run control system. Further improvements, enhancements, and implementations have already been planned, including the implementation of the agent reasoning engine based on JESS [6].

## Acknowledgments

The authors would like to acknowledge the prompt and responsive help provided by the Jade and Protégé-2000 development teams and communities.